\documentclass[a4paper]{jpconf}%
\usepackage{graphicx}
\usepackage{amsmath}
\usepackage{amsfonts}
\usepackage{amssymb}%
\providecommand{\U}[1]{\protect\rule{.1in}{.1in}}

\begin{document}

\title{Coarse Graining in Hydrodynamics and Effects of Fluctuations}
\author{T. Kodama and T. Koide}

\begin{abstract}
We address the physical meaning of hydrodynamic approach related with the
coarse graining scale in the frame work of variational formulation. We point
out that the local thermal equilibrium does not necessarily play a critical
role in the description of the collective flow patterns. We further show that
the effect of viscosity is also formlated in the form of the variational
method including fluctuations.

\end{abstract}

\address{Instituto de F\'{\i}sica, Universidade Federal do Rio de Janeiro, C.P.
68528, 21941-972, Rio de Janeiro, Brazil}

\ead{tkodama@if.ufrj.br, koide@if.ufrj.br}

\section{Introduction}

Hydrodynamic approach on event-by-event (EBE) basis has shown to be very
successful to describe the global and collective features of the data from
relativistic heavy-ion collisions, particularly of the behavior of the Fourier
components of flow pattern $\left\{  v_{n}\right\}  $ as a function of
centrality and transverse momenta data \cite{ebeHydro}. Such a success of the
hydrodynamic approach leads to the expectation that we can determine from a
detailed hydrodynamic analysis of experimental data, the initial dynamics of
the collisions together with the properties of QCD matter created, such as the
equation of state (EoS) and transport coefficients. On the other hand, these
successes brought us several new interesting questions and mysteries. The most
crucial one is why at all the hydrodynamic approaches work so well for such
violent and almost microscopic collisional processes. It is commonly believed
that the fundamental hypothesis for the validity of hydrodynamics is the local
thermal equilibrium (LTE). If this is true, and the hydrodynamic evolution is
the unique scenario for the description of the collision dynamics, then we are
led to conclude that the thermalization time and correlation length should be
extremely small. These would imply a very important consequence for the
further understandings of the QCD dynamics at extreme conditions. For AA
collisions this could still be acceptable, but surprisingly, the recent ALICE
experiment reports that the behavior of collective flow in the pA data seems
to be almost the same as that of AA collisions \cite{pA}. This casts a very
serious question for the proper physical meaning of hydrodynamic description
in pA collisions or, even in AA collisions \cite{Miklos}.

\section{Relativistic Hydrodynamics and Role of Coarse Graining}

Let us denote the conserved four-current density by $n^{\mu}(x),$ satisfying
the continuity equation, $\partial_{\mu}n^{\mu}(x)=0$. We consider the
Minkowsky coordinate, $x=\{x^{0},x\}$ with metric the $g_{\mu\nu
}=\operatorname{diag}\{1,-1,-1,-1\}$. The energy-momentum tensor $T^{\mu\nu
}(x)$ of the system also conserves, $\partial_{\mu}T^{\mu\nu}(x)=0$. Although
these 4 equations are far enough to determine the time evolution of these
quantities, but in some special physical situations, the total number of
variables reduces drastically. Suppose that, in the Landau frame (that is, the
energy flow rest frame), the spatial part of $T^{\mu\nu}(x)$ becomes isotropic
for any $x$. Furthermore, the rest frames for the matter and energy current
coinside. In such a situation, by introducing the Equation of State (EoS)
which establishes a functional relation among the local quantities,
$\varepsilon,P$ and $n$, as $P=P\left(  \varepsilon,n\right)  $, we obtain the
closed set of equations. This is called the ideal hydrodynamics and the
explicit forms of the time evolution equations are, $\ \partial_{t}%
\varepsilon+\left(  \vec{v}\cdot\nabla\right)  \varepsilon=-\gamma^{-1}\left(
\varepsilon+P\right)  \theta,\ \ \ \partial_{t}n+\left(  \vec{v}\cdot
\nabla\right)  n=-\gamma^{-1}n\theta,\ $and $\ d\left\{  \left(
\varepsilon+P\right)  /n\ \vec{u}\right\}  /dt=-\nabla P/n^{\ast}$ where
$n^{\mu}$ is the time-like eigenvector of $T^{\mu\nu}\left(  x\right)  $ with
$\theta=\partial_{\mu}u^{\mu}$, $\gamma=u^{0}$, $\vec{v}=\vec{u}/\gamma$ and
$n^{\star}=\gamma n$.

In the above, the assumption that EoS $P=P\left(  \varepsilon,n\right)  $ is
locally satisfied in the strict sense is somewhat a very severe condition. For
example, when we consider the hydrodynamic description of the relativistic
heavy-ion collisions, the size of the typical fluid element cannot be taken
too much smaller than that of the whole system, otherwise the degree of
freedom contained in the fluid element becomes too little for any
thermodynamical quantities to be defined. Furthermore, the time scale of the
collective motion cannot be much larger than the microscopic one because of
the very rapid expansion in relativistic heavy-ion collisions. Then, a given
fluid element suffers from large fluctuations and inhomogeneity in terms of
the microscopic configurations and it will be difficult to ignore the
deviation from the thermodynamic limit in EoS. However, we will argue in below
that the hydrodynamic responses in, for example, the collective flow do not
necessarily require the strictly local EoS in the EBE basis.

For example, let us consider a classical microscopic system which contains a
large number of quickly moving point-like particles. Then, the density
$n_{0}^{\ast}$ is a sum of the Dirac delta functions. However, we usually do
not require a very precise resolution both in space and in time to describe
the collective flow behaviors. Thus we introduce an averaged smooth density
distribution $\tilde{n}^{\ast}\left(  \vec{x},t\right)  $ from the original
distribution $n_{0}^{\ast}\left(  x\right)  $ using a 4 dimensional smoothing
kernel with limited support\cite{EuroJ} $W\left(  x\right)  =U_{\tau}\left(
t\right)  \times W_{h}\left(  \vec{x}\right)  $
\begin{equation}
\tilde{n}^{\ast}\left(  \vec{x},t\right)  =\int dt^{\prime}\int d^{3}\vec
{X}\ n_{0}^{\ast}\left(  \vec{X}\right)  \ U_{\tau}\left(  t^{\prime
}-t\right)  W_{h}\left(  \left\vert \vec{x}-\vec{r}\left(  t^{\prime};\vec
{X}\right)  \right\vert \right)  , \label{nsmooth}%
\end{equation}
Typically $U$ and $W$ are given by the Gaussian distributions with,
respectively, width $\tau$ and $h$, which characterize the scales of the time
and space resolutions. Similarly, the smoothed spatial current vector
$\mathbf{\tilde{j}}\left(  \vec{x},t\right)  $ can be defined, satisfying the
continuity equation, $\partial\tilde{n}^{\ast}/\partial t+\nabla_{x}%
\cdot\mathbf{\tilde{j}}=0$.

Using these current and density, the four-current, $\tilde{j}^{\mu}=\left(
\tilde{n}^{\ast},\mathbf{\tilde{j}}\right)  $ and the proper density,
$\tilde{n}=\sqrt{\tilde{j}^{\nu}\tilde{j}_{\nu}}$ can be composed. The
smoothed four-velocity field is then defined as $\tilde{u}^{\mu}=\ \tilde
{j}^{\mu}/n$. On the other hand, the smoothed energy-momentum tensor
$\tilde{T}^{\mu\nu}$ can be introduced in analogous way as the convolution of
the original $T^{\mu\nu}$ using the same smoothing kernel. Such
energy-momentum tensor again satisfies the continuity equation, $\partial
_{\mu}\tilde{T}^{\mu\nu}=0$. From this smoothed energy-momentum tensor, we can
calculate the smoothed proper energy density as $\tilde{\varepsilon}%
\equiv\tilde{u}_{\mu}\tilde{u}_{\nu}\tilde{T}^{\mu\nu}$. The smoothed proper
energy density defined in this way is an average of the energy density
observed in the rest frame of the \textit{matter} flow. The average is taken
over all contributions within the range of the coarse-graining scale in
space-time. In terms of the hydrodynamic modeling, we take these smoothed
quantities as the dynamical variables to represent the coarse-grained
system\cite{EuroJ}.

Let us consider one collision event which is characterized by a microscopic
state. Then we can calculate these hydrodynamic variables of this state
following the method described above. However, it is obvious that there exist
many different microscopic configurations which give the same hydrodynamic
response. Let us prepare the set of collision events described by microscopic
configurations which gives a given four-current $\tilde{j}^{\mu}$ at the
initial time $t_{0}$, and call this set $\Omega$. If we calculate
$\tilde{\varepsilon}$ at a space-time point $x$ for each event in $\Omega$,
each value of $\tilde{\varepsilon}$ is not same in general. This is true even
for the time evolution, $\tilde{j}^{\mu}\left(  \vec{x},t\right)  $ with
$t>t_{0}$.

However, if the coarse-graining size is increased, the number of the
microscopic configuration in the ensemble $\Omega$ at the point $x$ becomes
sufficiently large so that $\tilde{\varepsilon}$ and $\tilde{n}$ distribute
sharply around their mean-values, $\mathbf{\tilde{\varepsilon}}$ and
$\mathbf{\tilde{n}}$, respectively, as a consequence of the central limit
theorem. Furthermore, since $\tilde{\varepsilon}$ and $\mathbf{\tilde{n}}$ are
the average energy and matter densities belonging to the same fluid element,
we expect that they have a strong correlation, in such a way that
$\mathbf{\tilde{\varepsilon}}$ can be expressed as a function of
$\mathbf{\tilde{n}},$ $\mathbf{\tilde{\varepsilon}}=\mathbf{\tilde
{\varepsilon}}\left(  \mathbf{\tilde{n}}\right)  $.

Suppose that the fluctuations in $\tilde{\varepsilon}$ and $\tilde{j}^{\mu}$
are not important in the way that the system is characterized basically by the
densities $\mathbf{\tilde{n}}^{\ast}$and $\mathbf{\tilde{\varepsilon}}$. In
such a case, we expect that the most promising dynamics will be determined by
the optimization of the model action,
\begin{equation}
I=-\int d^{4}x\ \mathbf{\tilde{\varepsilon}}\left(  \frac{1}{\gamma
}\mathbf{\tilde{n}}^{\ast}\right)  . \label{ModelAction}%
\end{equation}
It is known that this procedure reproduces the model of the ideal
hydrodynamics \cite{Hydro-action}. As a result, the dynamics of the system
belonging to a given $\Omega$ is described by the hydrodynamic model as the
consequence of the coarse graining. There, the realization of LTE is not
required for real EBE basis. When the effect of the fluctuations in
$\tilde{\varepsilon}$ and $\tilde{j}^{\mu}$ plays an important role, see Sec. 4.

\section{\noindent Necessity of Real Event by Event Analysis}

As shown above, the hydrodynamic description in heavy-ion collisions reduces
to a coarse-grained dynamics obtained by the optimization of the model action
(\ref{ModelAction}) under the assumption of the existence of an effective EoS,
$\mathbf{\tilde{\varepsilon}=\tilde{\varepsilon}}\left(  \mathbf{\tilde{n}%
}\right)  $. Therefore, the success of ideal hydrodynamic modeling of
relativistic heavy ion collisions depends on the consistent choice of the
assumed EoS and the model action. These two conditions will be satisfied for a
broader range of microscopic configurations than that required by the real
"local thermal equilibrium".

On the other hand, the size of $\Omega$ depends on the coarse-graining scale.
For larger $\Omega$, the two conditions have a better chance to be satisfied.
We however loose the better resolution in the space-time recognition for
larger coarse graining size. In fact, we cannot observe the inhomogeneities
with smaller wavelength than the coarse-graining scale. This affects directly
the class of observables that the model can describe. Even though some
observables might be insensitive to the inhomogeneities in each event. As an
extreme example, we take the situation where the coarse-graining size is
larger than the system size and total time evolution. Then the ensemble
$\Omega$ can be regarded as the statistical ensemble of the whole system
itself, and the resultant system is a simple fire-ball model. The thermal
model for the particle ratio can be considered in this category.

The coarse-graining size is thus intimately related to the class of
observables and the validity of hydrodynamic description. For some observables
which do not require a precise space-time resolution, the real EBE
hydrodynamics with LTE is not necessary and the effective hydrodynamic
description for the statistical ensemble $\Omega$ will be sufficient for the
understanding of the physics of these observables. As a matter of fact, the
experimental observables are usually averaged over collision events classified
in terms of the initial configuration rather loosely defined, such as
centrality, event plane, etc. In other words, the present collective flow data
are still of inclusive nature. In order to claim that the real hydrodynamics
with LTE is valid, we need to have observables that reflect the genuine
hydrodynamic profile in EBE basis. For example, the remnant of a sharp shock
wave propagation, if exists, would be a good evidence and it also tells the
possible coarse-graining size of the collective flow, since a shock wave is a
genuine local hydrodynamic phenomena. The shock thickness should not be larger
than the coarse-graining scale of the collective flow.

The key point is that when we apply the hydrodynamic modeling, we do not know
a priori the coarse-graining scale suitable for the flow variables in the real
scenario. This puts a certain limitation in extracting the meaningful
information of the initial condition. For this purpose, it is essential to
find out the set of observables which carry the information on the
inhomogeneities of the initial conditions on the EBE basis. The flow
parameters $\left\{  v_{n}\right\}  $, often called "event-by-event" analysis,
in the sense that correlations among different observables measured for each
event in coincidence, but there still exists a huge statistical ensemble which
gives the same observed correlation. For example, the cumulant method to
determine the flow parameters eliminates the information of event plane. In
the recent paper, it is pointed out that event plane may differ in low and
high $p_{T}$ domain \cite{EuroJ}, according to the coarse graining scale. If
it can be experimentally measured, it would furnish some information on coarse
graining scale in heavy-ion collisions.

\section{Fluctuation of Fluid Variables and Stochastic Variational Method}

Within the vision that hydrodynamic evolution is an effective dynamics for
coarse-grained variables for the energy-momentum tensor, each real collisional
event is an element of the statistical ensemble $\Omega$ and does not obey a
unique time evolution equation due to the difference in the microscopic
degrees of freedom to which our macroscopic hydrodynamic variables are blind.
When the fluctuation of events in $\Omega$ is large, they should be taken into
account in the determination of dynamics of coarse-grained hydrodynamic
variables. Then the variation procedure in Eq. (\ref{ModelAction}) should be
modified so as to include the effect of the fluctuation which was ignored in
Sec. 3. The stochastic variational method (SVM) is known as an appropriate
approach for such situations\cite{koide}.

In SVM, we have to introduce two stochastic differential equations (SDE), one
for the forward direction in time (FSDE), $d\mathbf{r}(t)=\mathbf{u}%
dt+\sqrt{2\nu}d\mathbf{W}(t),~~(dt>0),\ \ $and the other, backward in time
(BSDE), which describes the time reversed process of FSDE, $d\mathbf{r}%
(t)=\tilde{\mathbf{u}}dt+\sqrt{2\nu}d\tilde{\mathbf{W}}(t),~~(dt<0),$ where
$\nu$ is the strength of the noise and $\mathbf{W}(t)$ and $\tilde{\mathbf{W}%
}$ are the independent Wiener processes. $\mathbf{u}$ and $\tilde{\mathbf{u}}$
are the velocity fields for the forward and backward SDEs, respectively. These
two SDEs are necessary to accommodate the fixed initial and final boundary
conditions in the variational procedure. Therefore, trajectories specified in
the two SDE should describe the same physical ensemble. To satisfy this
condition, the two Fokker-Planck equations which are derived from FSDE and
BSDE should be equivalent. This leads to the consistency condition,
$\mathbf{u}=\tilde{\mathbf{u}}+2\nu\nabla\ln\rho,$where $\rho$ is the particle
density which is given by the solution of the Fokker-Planck equation. SVM
proposes to determine these velocity fields from an action through the
variation principle.

We started from the classical action which leads to the ideal Euler equation
for a non-relativistic fluid,
\begin{equation}
I(\mathbf{r})=\int_{t_{a}}^{t_{b}}dt\int d^{3}\mathbf{R}\left[  \frac{\rho
_{0}^{m}(\mathbf{R})}{2}\left(  \frac{d\mathbf{r}(\mathbf{R},t)}{dt}\right)
^{2}-J\varepsilon\right]  ,
\end{equation}
where $\mathbf{r}(\mathbf{R},t)$ is the Lagrangian coordinate associated with
the fluid element $\mathbf{R,}$ and $\varepsilon$ is an internal energy
density, $\mathbf{J}$ is the Jacobian between $\mathbf{r}$ and $\mathbf{R}$,
and the mass density is defined by $\rho_{m}=m\rho$. The spatial integral is
done over the initial position $\mathbf{R}$ of fluid elements, weighted by the
mass density $\rho_{0}^{m}(\mathbf{R})$. The corresponding stochastic
Lagrangian density is given by
\begin{equation}
\mathcal{L}=\frac{\rho_{0}^{m}}{2}\left[  \left(  \frac{1}{2}+\alpha
_{2}\right)  \left\{  \left(  \frac{1}{2}+\alpha_{1}\right)  \mathbf{u}%
^{2}+\left(  \frac{1}{2}-\alpha_{1}\right)  \tilde{\mathbf{u}}^{2}\right\}
+\left(  \frac{1}{2}-\alpha_{2}\right)  \tilde{\mathbf{u}}\cdot\mathbf{u}%
\right]  -J\varepsilon. \label{L}%
\end{equation}
Here $\alpha_{1}$ and $\alpha_{2}$ is arbitrary constants, come from the
ambiguity for the stochastic representation of the kinetic term. The
corresponding action is an average over the whole SDE solutions, and the
variation is taken with respect to the unknown fields, $\mathbf{u}$ and
$\tilde{\mathbf{u}},$ with the constraints that they coincide with the
so-called conditional average velocity fields, $D\mathbf{r}$ and
$D\mathbf{\tilde{r}}$.

After applying SVM, we obtain
\[
\rho^{m}\left(  \partial_{t}+\mathbf{u}_{m}\cdot\nabla\right)  u_{m}^{i}%
-\sum_{j}\partial_{j}(\mu\rho_{m}e_{ij}^{m})-2\kappa\rho^{m}\sum_{j}%
\partial_{i}(\sqrt{\rho^{m}}^{-1}\partial_{j}^{2}\sqrt{\rho^{m}})=-\nabla P,
\]
where $\eta=\alpha_{1}(1+2\alpha_{2})\nu\rho^{m}$, $\kappa=2\alpha_{2}\nu^{2}$
and $e_{ij}^{m}=\partial_{j}u_{m}^{i}+\partial_{i}u_{m}^{j}$. The pressure $P$
is defined by $(\rho_{m})^{2}d(\varepsilon/\rho_{m})/d\rho_{m}$. One can see
that the second term on the left hand side corresponds to the viscosity and
this equation is reduced to the Navier-Stokes (NS) equation when we set
$\alpha_{2}=0$. That is, the fluctuation effects which were ignored in Sec. 3
induces the effects of viscosity in accordance with the
fluctuation-dissipation theorem\footnote{Exactly speaking, the second
coefficient of viscosity vanishes in this equation. To obtain a finite value,
we have to consider the variation of entropy.}. For other results of different
values of $\alpha^{\prime}s$, see Ref. \cite{koide}.

\section{Concluding remarks}

In this work, we studied some fundamental questions of hydrodynamic approach
to the description of relativistic heavy ion collisions, in particular, the
validity and meaning of LTE. We introduced explicitly the coarse-graining
procedure for the hydrodynamic modeling together with its variational
formulation. In this picture, the collective flow patterns can be reproduced
without requiring the LTE in a strict sense. That is, the hydrodynamic
behavior observed in relativistic heavy-ion collisions does not necessarily
imply the realization of LTE \footnote{This has been already noticed before.
See for example, Ref. \cite{Larry}}. We further discussed possible signals for
coarse graining scale and genuine hydrodynamic behaviors on event by event
basis. For example, the remnant of a sharp shock wave propagation would be a
good observable which tells the possible coarse-graining size of the
collective flow. Another example is to determine the event plane for different
transverse momentum domain. Finally we showed that the coarse-graining is
intimately related to the origin of viscosity and this effect can be
formulated in the variational method extending dynamical variables to
stochastic domain. In order to quantify the questions raised here, it will be
useful to perform the analysis of coarse-graining described in this work for a
certain microscopic model which gives complete dynamical evolution of the
energy-momentum tensor, such as PHSD\cite{Elena}.

\smallskip

This work is supported by CNPq, FAPERJ and PRONEX of Brazil, and also a part
of the work has been supported by EMMI-Helmholtz Institute, Germany.

\end{document}